\documentclass[%
 reprint,
superscriptaddress,
 amsmath,amssymb,
 aps,
prb,
]{revtex4-2}

\usepackage{graphicx}% Include figure files
\usepackage{bm}% bold math
\usepackage{hyperref}% add hypertext capabilities
\usepackage[mathlines]{lineno}% Enable numbering of text and display math

\usepackage{siunitx}

\makeatletter
\def\maketitle{
\@author@finish
\title@column\titleblock@produce
\suppressfloats[t]}
\makeatother

\begin{document}

\title{Elucidating magnetic structure with optical dopants: erbium-doped Gd\texorpdfstring{\textsubscript{2}}{₂}SiO\texorpdfstring{\textsubscript{5}}{₅}}

\author{Luke S. Trainor}

\author{Masaya Hiraishi}

\affiliation{Department of Physics, University of Otago, Dunedin New Zealand}
\affiliation{Dodd-Walls Centre for Photonic and Quantum Technologies, Dunedin, New Zealand}

\author{J.-R. Soh}
\affiliation{Quantum Innovation Centre (Q.InC), Agency for Science, Technology and Research (A*STAR), 2 Fusionopolis Way, Innovis \#08-03, Singapore 138634, Singapore}
\affiliation{Centre for Quantum Technologies, National University of Singapore, 3 Science Drive 2, Singapore 117543, Singapore}

\author{Jevon J. Longdell}
\affiliation{Department of Physics, University of Otago, Dunedin New Zealand}
\affiliation{Dodd-Walls Centre for Photonic and Quantum Technologies, Dunedin, New Zealand}

\date{\today}

\begin{abstract}
The narrowness of the optical transitions of rare-earth-ion dopants makes them highly sensitive probes of their environment.
We measured the optical transitions Er\textsuperscript{3+} dopants to determine the previously unknown magnetic ordering of Gd\textsubscript{2}SiO\textsubscript{5} -- a promising host for quantum applications of rare-earth dopants. 
By measuring the  transitions' magnetic-field dependence
we determined an antiferromagnetic ordering 
with spins oriented along or slightly canted from the crystal's $a^*$ axis.
The optical transitions are narrower than the coupling to gadolinium spins revealing information about the coupling strengths.
We further optically measured a N{\'e}el temperature of \qty[parse-numbers=false]{1.86\pm0.01_\mathrm{stat.}\pm0.07_\mathrm{syst.}}{\kelvin}, and assembled a phase diagram in applied field and temperature showcasing a triple point where two gadolinium sites order semi-independently from each other.
At high applied field the erbium dopants show long optical coherence times up to \qty{0.4}{\milli\second} at \qty{3}{\tesla};
at low fields these are probably limited by three low-frequency magnon modes below \qty{10}{\GHz}, observed directly.
This study can be used to benchmark a method of magnetic structure determination.  
\end{abstract}

\maketitle

\section{\label{sec:intro} Introduction}

Determining the order of magnetic quantum materials is vital for designing devices which harness their unique characteristics, as well as engineering new materials.
A variety of indirect techniques are commonly used to determine magnetic ordering, such as measurements of heat capacity and magnetic susceptibility.
The gold-standard method is neutron scattering as it directly probes the magnetic order: the magnetic dipole moments of the neutrons diffract off the magnetic configuration in the crystal~\cite{Boothroyd_book}.
Some materials, however, are difficult to measure with neutron scattering as their component elements have large neutron-absorption cross sections. \textsuperscript{157}Gd in particular has the largest thermal neutron-absorption cross section of any stable isotope \cite{searsNeutronScatteringLengths1992}.
Therefore to study gadolinium-containing compounds, alternative techniques such as resonant x-ray scattering \cite{detlefsMagneticStructureGdNi$_2$B$_2$C1996} are used, as well as highly specialized hot neutron beams \cite{palaciosMagneticStructureMagnetocalorics2014}, which are absorbed less readily.

We present a different approach to magnetic structure determination: rare-earth ion dopants. Many trivalent rare-earths have large magnetic moments and characteristic, narrow optical transitions \cite{kukharchykOpticalCoherence2018}.
We report an optical spectroscopic study of erbium-doped gadolinium oxyorthosilicate (Er\textsuperscript{3+}:Gd\textsubscript{2}SiO\textsubscript{5}, or Er:GSO) under various temperatures and applied magnetic fields. As far as we are aware, the magnetic ordering of the host material has not been experimentally studied before \cite{yangStrongMagnetocaloricCoupling2023}. Therefore, through measuring optical transitions of erbium dopants in GSO, we not only characterize its potential as a rare-earth host material, but also demonstrate that optical spectroscopy can be used to determine fundamental magnetic properties such as magnetic structure and phase transition temperatures and fields. 

Our interest in Er:GSO comes from its potential as a magnetically quiet host for rare-earth quantum technologies.
Coherently manipulating optical and spin states in rare-earth ions doped in crystals enables them to be used for various applications such as quantum memories \cite{afzeliusMultimodeQuantumMemory2009}, and microwave--optical transduction \cite{xieScalableMicrowavetoopticalTransducers2025}. Their 4$f$ electrons are electrostatically shielded by outer orbitals, giving narrow optical \cite{kukharchykOpticalCoherence2018} and spin transitions \cite{probstAnisotropicRareEarthSpin2013} with long coherence times, even in crystals \cite{ledantecTwentythreemillisecondElectronSpin2021}. 
In particular, yttrium orthosilicate (Y\textsubscript{2}SiO\textsubscript{5}, or YSO) -- a non-magnetic insulator -- stands out as one of the most commonly used rare-earth host crystals for quantum applications \cite{lago-riveraTelecomheraldedEntanglementMultimode2021,fernandez-gonzalvoCavityenhancedRamanHeterodyne2019,jiangQuantumStorageEntangled2023}, due to its low nuclear spin density, which leads to small magnetic fluctuations at dopant rare-earth ions \cite{pignolDecoherenceInducedDipoledipole2024}, allowing for high coherence \cite{boettgerMaterialOptimizationEr32003,equallUltraslowOpticalDephasing1994,chiossiOpticalCoherenceSpin2024,zhongOpticallyAddressableNuclear2015,rancicCoherenceTimeSecond2018}.

Recent studies showed that rare-earth doped magnetically ordered crystals are another candidate for those quantum applications. Previously such materials had been largely ignored due to their high spin density; flips of unpaired electronic spins in crystals are a particularly bad decoherence source in rare-earth crystals due to their much larger dipole moment than nuclei. However, such magnetic fluctuations can be suppressed by magnetic ordering. Long optical coherence times were reported in erbium-doped antiferromagnetic gadolinium vanadate (Er:GdVO\textsubscript{4}) \cite{hiraishiLongOpticalCoherence2025c}, comparable to those seen in erbium-doped yttrium vanadate (Er:YVO\textsubscript{4}) \cite{liOpticalSpectroscopyCoherent2020b}, a non-magnetic counterpart.

Magnetically ordered crystals also support magnons -- collective excitations of spins. In antiferromagnetic rare-earth crystals, ordering temperatures $O(\qty{1}{\K})$ are common, giving magnons of frequency $O(\qty{1}{\K}\cdot k_B/h\approx\qty{20}{\GHz})$. 
This magnon frequency is similar to superconducting qubit frequencies for which efficient microwave--optical transduction is sought \cite{hanMicrowaveopticalQuantumFrequency2021,rochmanMicrowavetoopticalTransductionErbium2023,forschMicrowavetoopticsConversionUsing2020,mirhosseiniSuperconductingQubitOptical2020}, leading to the proposal that bulk magnons coupled to rare-earth optical transitions could provide efficient transduction \cite{Everts_2019,puelEnhancementMicrowaveOptical2025}.

From the popularity of YSO, and recent success in rare-earth magnetic materials, it seems natural to investigate GSO, a magnetic counterpart of YSO with a lower nuclear-spin background \cite{kanaiUnifiedPeriodicTable2026}.
However, we must here point out that GSO crystallizes in a different structure to the most common form of YSO.
The most commonly used form of YSO is known as X\textsubscript{2}-YSO ($C2/c$ space group), whereas GSO crystallizes into the same structure as X\textsubscript{1}-YSO ($P2_1/c$ space group) \cite{becerroRevisionCrystallographicData2004}.
X\textsubscript{1}-YSO has been studied much less in the literature \cite{askariLatticeDynamicsDielectric2022}, mainly because it requires low-temperature growth, which is more suited to nanoparticle growth methods such as the sol--gel method \cite{wangPreparationXrayCharacterization2001}.

This study therefore measures the properties of the host material and erbium dopants in parallel, and finds an antiferromagnetic host material that acts as a very quiet dopant environment at large applied fields.

\section{Results}

\begin{figure}
    \centering
    \includegraphics[width=\linewidth]{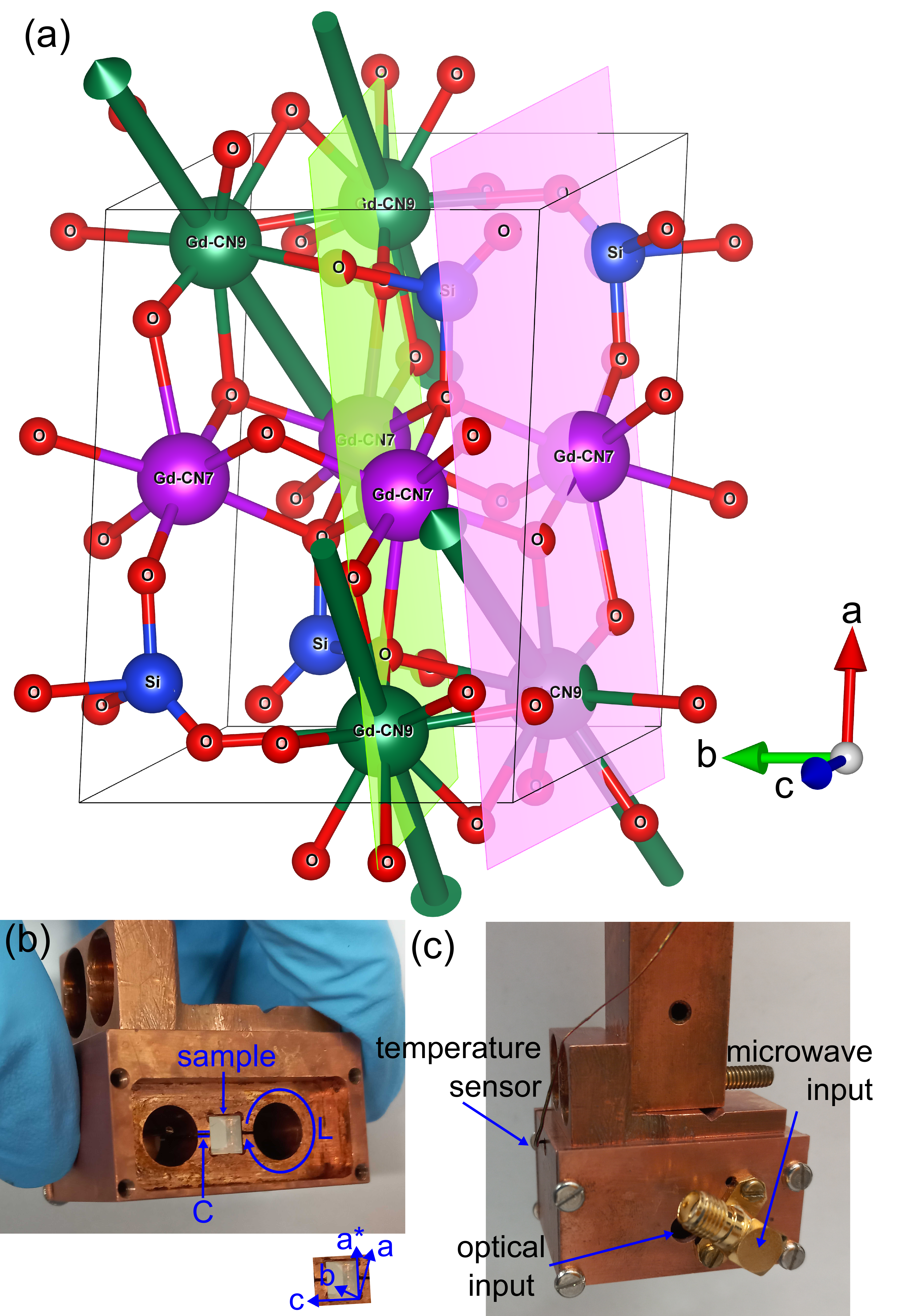}
    \caption{(a) Monoclinic unit cell of a GSO crystal \cite{dramicaninLuminescenceStructuralProperties2006}. Gadolinium ions are labelled with their oxygen coordination number, CN7 (purple) or CN9 (dark green). CN7 ions lie approximately on (100) sheets and have approximate $C_s$ ligand symmetry with mirror plane shown for one of the ions in pink. CN9 ions have approximate $C_{3v}$ ligand symmetry. The $C_3$ axes are shown with arrows and one of the mirror planes for a CN9 ion is shown in lime green. Erbium ions can substitute any site, and within the unit cell there are four subsites with differently oriented crystal fields. (b) A photograph of the sample inside a bulk loop-gap cavity with the front plate removed. The sample is annotated as well as a loop in the cavity giving an inductance $L$, and a gap giving a capacitance $C$, which makes the cavity an analog to a $LC$-resonator. Below: diagram of crystal axes in this sample. The $a^*$ axis is within \qty{0.5}{\degree} of the normal of the top and bottom faces, whereas the other axes are only approximate. The sign of the $b$ axis is unknown.
    (c) Closed photograph of the cavity without microwave cables.
    }
    \label{fig:sample}
\end{figure}

The centrosymmetric monoclinic unit cell of GSO is shown in Fig.~\ref{fig:sample}(a). It contains gadolinium ions residing at two different $4e$ Wyckoff sites with oxygen coordination number (CN) 7 and 9 respectively. Each site has four ions per unit cell, given the four-fold multiplicity of the Wyckoff site. The 49\,at.\,ppm erbium dopants can substitute at any gadolinium position.
Although the gadolinium sites have $C_1$ point symmetry, they are often ascribed approximate $C_s$ (CN7) and $C_{3v}$ (CN9) symmetries in the literature, referencing the positions of the coordinate oxygen ions \cite{lisieckiOpticalSpectraLuminescence2010,camargoSpectroscopicCharacteristicsEr32002}.
Our sample is mounted in a microwave cavity with optical access, shown in Fig.~\ref{fig:sample}(b,c). This cavity sits in the bore of a 3-T superconducting magnet, which can apply field vertically, along the crystal's $a^*$ axis ($b\perp a^*\perp c$).

\subsection{Optical Absorption Spectroscopy}

At \qty{4}{\kelvin} we observe signatures of erbium ions residing in two distinct gadolinium sites with transitions near \qty{1529}{\nm}: site 1 (\qty{195.99}{\THz}) and site 2 (\qty{196.09}{\THz}), which we assigned using the labelling of \textcite{camargoSpectroscopicCharacteristicsEr32002} by an emission-resolved photoluminescence measurement (see Methods). 
These transitions are from the ground doublet $Z_1$, which is the lowest energy doublet of the ${}^4I_{15/2}$ manifold to $Y_1$, the lowest doublet of ${}^4I_{13/2}$.
The closeness in frequency of the $Z_1\rightarrow Y_1$ transitions for the two sites is essentially a coincidence; other transitions are separated by much larger frequencies.
It is not known which of the two sites has oxygen coordination number (CN) of 7 or 9; it was argued based on ${}^4S_{3/2}$ lifetimes that site 1 may have CN7 \cite{camargoSpectroscopicCharacteristicsEr32002}, however our measurements of the ${}^4I_{13/2}$ lifetimes (see Methods) and magnetic-field dependence of the transitions would lead to the opposite conclusion; this is revisited later.

\begin{figure*}
    \centering
    \includegraphics[width=\linewidth]{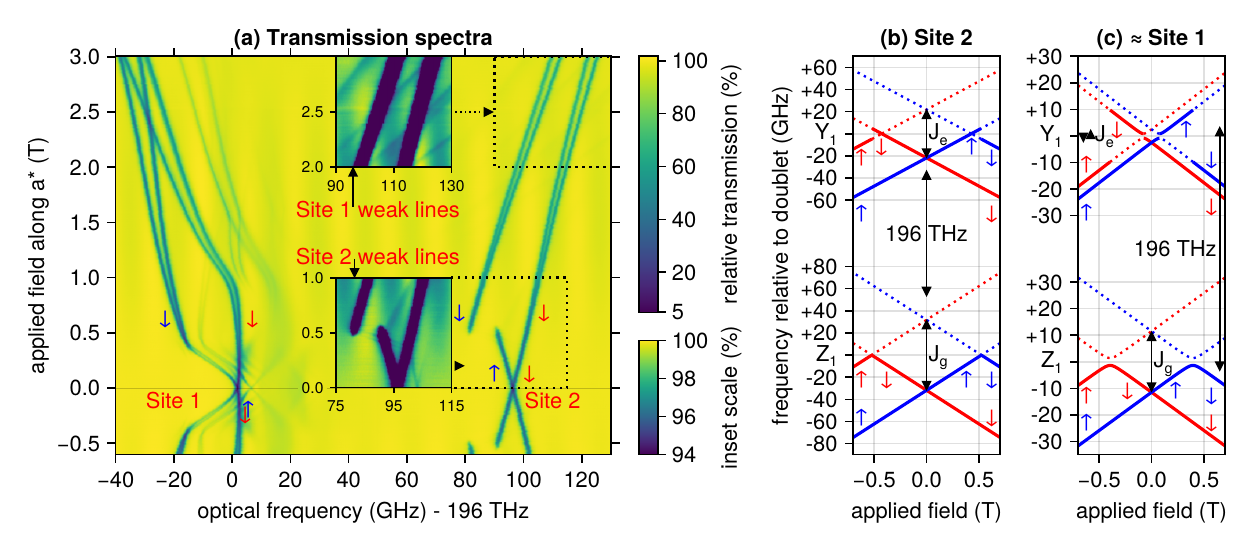}
    \caption{(a) Optical transmission spectra of Er:GSO at $\approx\qty{60}{\milli\K}$. Site 1 is seen around \qty{196}{\THz}, whereas site 2 is around \qty{196.1}{\THz}. Blue and red arrows next to strong absorption features label like-to-like transitions of antiferromagnetic sublattices and the erbium spin state of that sublattice at that field. Blue (red) arrows label the ions that are spin up (down) at zero applied field.  Top inset: weak transitions from site 1 observed behind site 2 lines. Bottom inset: A pair of weak transitions of site 2 seen around \qty{0.4}{\tesla}. Weak lines at other fields are possibly satellite lines or due to cooperative excitation of neighbouring gadolinium spins. 
    (b) Energy levels of site 2. Red and blue are magnetic sublattices that order antiferromagnetically, pointing along the $a^*$ axis. In the $Z_1$ ground doublet, the bold lines are the ground states and their spin orientation is annotated. Dotted lines are the unpopulated state. The strong transitions are like-to-like transitions that preserve spin direction, hence the excited $Y_1$ state that is observed in those transitions is bold, whereas the other is dotted. At about \qty{0.52}{\tesla}, the ground state of one of the subsites changes spin direction as the Zeeman splitting goes through zero and therefore we measure a different excited state. The $g$-factors are found from the strong transitions as well as the weak lines shown in the bottom inset of (a). (c) Approximate model for site 1 energy levels. The $g$-tensor does not have a principal axis along $a^*$ and/or the spins are canted from $a^*$, hence like-to-unlike transitions are stronger and energy levels do not cross.
    }
    \label{fig:coldspectra}
\end{figure*}

At the dilution refrigerator's base temperature of $\approx\qty{60}{\milli\kelvin}$, in the gadolinium magnetically ordered phase, the transition frequencies have drastically different applied-field dependence, 
see Fig.~\ref{fig:coldspectra}(a).
We expect that in the regime of low applied field, dopant erbium spins will order the same way as host gadolinium spins (via superexchange interaction), however it is plausible that erbium ions will cant due to their anisotropic $g$-tensors.
We do not observe hysteresis in the optical spectra, which leads us to believe the ordering is antiferromagnetic; each site will have spin-up and spin-down sublattices.

Examining site 2 in Fig.~\ref{fig:coldspectra}(a), we see linear Zeeman shifts in the low field regime, so we conclude it must order oriented along the applied-field direction ($a^*$ axis).
At zero applied field it has an absorption depth full width at half maximum (FWHM) of \qty{1.12}{\GHz} found from fitting the absorption to that of a Voigt profile; the tails of the inhomogeneous broadening have a decay in between Gaussian and Lorentzian.
Increasing applied field, we see the line splits into two as the spin-up and spin-down antiferromagnetic sublattices decrease and increase their Zeeman splitting respectively.
At \qty{0.5}{\tesla}, there is a discontinuity in one of the site 2 lines where the erbium ground state of the formerly spin-up sublattice flips to spin down. This change happens only to the erbium dopants, whereas the gadolinium ions in the same positions are unaffected.
Above this field we observe the same transition on both sublattices, but with an offset due to the exchange coupling to the differently aligned neighbour gadolinium ions. 
The energy levels of site 2 are illustrated in Fig.~\ref{fig:coldspectra}(b). 
At higher temperatures -- instead of a discrete jump -- one line fades in as the other fades out due to thermal population of the excited spin state.
Increasing the applied field beyond about \qty{0.8}{\T}, site 2 splits into four sublattices due to its interaction with site 1, which we will evidence later.

The low-field spectra of site 2 are readily explained by an effective spin-{\textonehalf} model for each doublet with mean exchange interaction along $a^*$ and insensitive $g$-factor about that direction, see Methods. 
The high degree of agreement with the model -- see Extended Data Figs.~\ref{fig:toysite2} and \ref{fig:toysitehot} at low fields and low canting of $\theta\approx\qty{0.5}{\degree}$ leads us to the conclusion that not only is the orientation along $a^*$, but that the $g$-tensors of the ground and excited states have the $a^*$ axis as a principal axis, or nearly so. 
This observation provides further evidence against the assignments of \textcite{camargoSpectroscopicCharacteristicsEr32002}, and leads us to believe that site 2 is CN7, as the CN9 site's $g$-tensor axes are likely set by its approximate $C_{3v}$ symmetry, which has a $C_3$ axis approximately \qty{45}{\degree} from the $a^*$ axis.

Site 1 shows a lot of structure.
At zero applied field, the strongest line has an Voigt profile absorption depth FWHM of \qty{1.22}{\GHz}. It splits immediately with applied field. The nearly field-insensitive line is the like-to-like transitions of spin-up and spin-down sublattices, whereas the lines that reduce in frequency to \qty{194985}{\GHz} at \qty{0.4}{\T} are a ``like-to-unlike'' transition, where the optically excited state has flipped spin compared to the ground state. 
The second strongest site-1 line at zero field is \emph{also} a like-to-unlike transition; the gap between them is due to a gap in the excited state energy levels arising from a small magnetic dipole moment perpendicular to $a^*$. 
Above about \qty{0.4}{\T}, there are four main lines with similar field dependence, and here all sublattices of erbium are approximately spin antialigned to the applied field.
An approximate energy-level diagram is shown in Fig.~\ref{fig:coldspectra}(c).
A number of weaker lines are observed around site 1, which we believe are transitions where an erbium optical transition occurs simultaneously to a gadolinium spin excitation \cite{hiraishiLongOpticalCoherence2025c,hiraishiOpticalSpectroscopySingle2025}.

Site 1 is explained qualitatively by an effective spin-\textonehalf{} model, but either a canting of $\theta\sim\qty{20}{\degree}$  between the mean field and applied field axis, or anisotropic $g$-tensors canted with respect to the $a^*$ axis are required. 
The CN9 site likely has a principal $g$-tensor from its approximate $C_3$ axis about $\sim\qty{45}{\degree}$ from the $a^*$ axis, which we have assumed along with $a^*$ orientation to make Fig.~\ref{fig:coldspectra}(c), and a comparison to the model is give in Extended Data Fig.~\ref{fig:toysite1C3v}.
We therefore conclude that site 1 may point along the $a^*$ axis or have canting angle up to $\sim\qty{20}{\degree}$ with respect to it.

Site 1 has a nearly degenerate ground-state Zeeman doublet near \qty{0.4}{\T}, which is evidenced by a weak thermal line about \qty{2}{\GHz} lower in frequency at that field.
Considering the site-1 effective spin-\textonehalf{} models have $g$-factors of $\sim4$ from the like-to-unlike transitions, compared to gadolinium's g-factor of $\approx2$, we might expect a change in ordering behaviour at about \qty{0.8}{\T}. Indeed, at about \qty{0.9}{\T}, two site-1 lines near \qty{196001}{\GHz} show a change in slope of their field dependence, which may be a gradual change in neighbouring site-1 gadolinium ordering at that field. Additionally, two weak lines emerge next to those strong transitions, plausibly showing that some nearby gadolinium have nearly zero Zeeman splitting and are canting. Additionally, site 2 shows marked splitting into four sublattices above this field.

We therefore conclude that site 2 orders antiferromagnetically, oriented along the $a^*$ axis, and that site 1 orders antiferromagnetically, oriented along or close to this axis. Thus there are four magnetic sublattices at zero field -- two in each site. The assigned order disagrees with the calculation of \cite{yangStrongMagnetocaloricCoupling2023} who found an order with two magnetic sublattices: CN7 and CN9 Gd\textsuperscript{3+} formed two ferromagnetically ordered sublattices, oriented along the $c$ axis, but the relative order between the two sublattices was antiferromagnetic.

\begin{figure*}
    \centering
    \includegraphics[width=\linewidth]{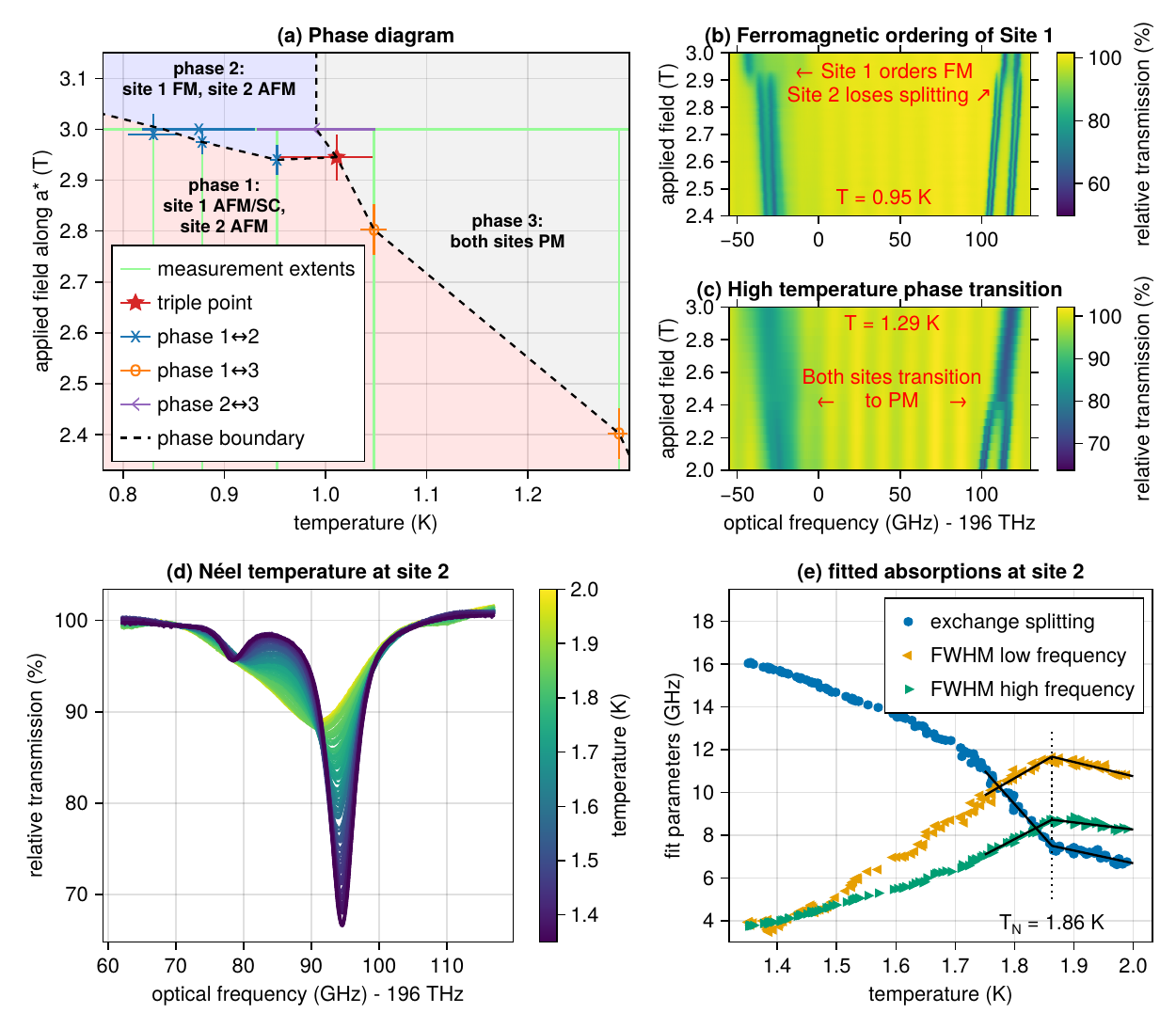}
    \caption{
    (a) A phase diagram assembled through optical measurements near a triple point. Light green lines in the background show the extent of our measurements. Phase transition fields were determined as sudden global changes on optical transmission spectra. Where the optical spectra show features from both phases, the range of coexistence in applied field is shown by error bars. Points with larger temperature uncertainty were measured by sweeping the sample's temperature instead. For this graph the systematic error from the temperature sensor is not included as it is expected to be consistent. At \qty{2.95}{\tesla} and \qty{1.011}{\kelvin}, the optical spectra had features from all three spectra, and this point is marked as the triple point. 
    Dashed lines and background colour are guides to the eye. FM: ferromagnetic; AFM: antiferromagnetic; PM: paragmagnetic; SC: spin canting.
    (b) Optical spectra at \qty{950\pm7}{\milli\kelvin} showing the phase $1\leftrightarrow2$ transition. The four main transitions of site 1 approach two degenerate pairs. At that field, a phase transition occurs and at higher applied field only one main transition is observed. We therefore believe that site 1 becomes ferromagnetically ordered. Site 2 goes from four transitions to two, so we believe that it is still ordered antiferromagnetically, but that the former two$\rightarrow$four splitting was due to nonequivalent site 1 neighbours. 
    (c) Phase-transition between phases $1\leftrightarrow3$ at \qty{1.29\pm0.01}{\kelvin}.
    (d) Optical transmission through the sample at site 2 as the sample is slowly cooled through and beyond the Néel temperature at zero applied field. (e) The spectra from (d) are fit to a sum of two Lorentzian line shapes. The splitting and full widths at half maximum (FWHM) are shown and are used to determine the Néel temperature where their temperature dependence changes slope suddenly. Solid black lines show the fitted parameters in the range over which they are fitted, and the dotted black line shows the extracted Néel temperature.
    }
    \label{fig:phasediagram}
\end{figure*}

The four sublattices of site 1 look like they will become two degenerate pairs at about \qty{3.4}{\tesla}. Given our prediction that they are canting in the high field range, degeneracy would indicate that they are nearing equal spin orientations. To investigate this, we took optical measurements at \qty{950}{\milli\kelvin}, shown in Fig.~\ref{fig:phasediagram}(b). At this temperature, the saturation field moves down to about \qty{2.97}{\tesla}. There the two pairs of sublattices become degenerate, which causes a phase transition, to a phase with ferromagnetic site 1. The number of observed lines in site 2 decreases from four to two, and indicates that the splitting that has disappeared was due to exchange coupling to magnetically nonequivalent site-1 gadolinium ions.

Fig.~\ref{fig:phasediagram}(a) plots the magnetic phase diagram of GSO, based on the optical spectra of erbium dopant ions as a proxy, including transitions to the high-temperature phase, exemplified in Fig.~\ref{fig:phasediagram}(c).

We additionally measured the Néel temperature by observing the optical transmission at site 2 as temperature was lowered gradually from \qty{2}{\K} to \qty{1.35}{\K}, shown in Fig.~\ref{fig:phasediagram}(d).
At the low end of this temperature range, the transmission shows two well defined absorption dips, whereas at higher temperatures, only one asymmetric dip is seen.
A change in the temperature dependence of the Zeeman splitting is known to signal a phase-transition in a range of magnetic materials \cite{leaskStatusMagnetoopticalExperiments1968,leaskMagnetoOpticalInvestigationTrivalent1969}.
To extract a Néel temperature from this measurement, the transmission is fit to an absorption assuming two Lorentzian absorbers: $T=\exp(-\alpha(\nu)L)$, where $\alpha(\nu)L = \sum_{i=1,2}A_i\gamma_i/((\nu-\nu_i)^2+\gamma_i^2)$ is the total absorption depth for two absorptions of central frequencies $\nu_i$, FWHMs $2\gamma_i$ and strength $A_i$. The temperature dependence of the exchange splitting $\lvert\nu_2-\nu_1\rvert$, and FWHMs are shown in Fig.~\ref{fig:phasediagram}(e). Those parameters show a change in slope at the Néel temperature, which is determined to be \qty[parse-numbers=false]{1.86\pm0.01_\mathrm{stat.}\pm0.07_\mathrm{syst.}}{\kelvin}. The statistical uncertainty is mainly in how we choose the function for a change in slope, but
the majority of the error in this value is systematic and comes from the thermistor.

\subsection{Microwave Spectroscopy}

\begin{figure}
    \centering
    \includegraphics[width=\linewidth]{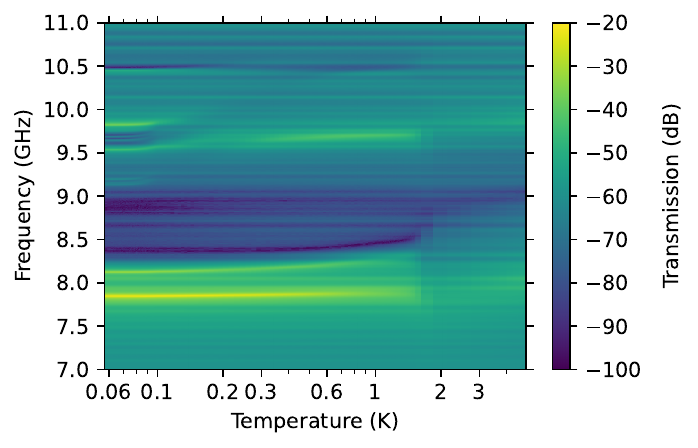}
    \caption{Microwave transmission as the dilution refrigerator is cooled from \qty{4}{\K} to base temperature. The N\'eel temperature is seen as a sudden reduction of mode linewidth. Below about \qty{100}{\milli\K} a cavity mode is split by strong coupling to a Gd\textsuperscript{3+} magnon at \qty{9.7}{\GHz}.}
    \label{fig:mwcooldown}
\end{figure}

Temperature dependence of the gadolinium resonances was investigated as the cryostat was cooled down from 4.9\,K to 57\,mK, shown in Fig.~\ref{fig:mwcooldown}.
Cavity modes between 8 and 10\,GHz show drastic change at about 1.72\,K. Below that temperature, they narrow significantly as Gd\textsuperscript{3+} magnon modes narrow in linewidth due to the magnetic phase transition \cite{evertsUltrastrongCouplingMicrowave2020}. The temperature is consistent with that determined by the optical measurements, noting that the optical measurements were performed with much greater temperature resolution.

At much lower temperatures, below 100\,mK, a single cavity mode at about \qty{9.7}{\GHz} splits into two. 
The split represents an avoided crossing due to coupling to a Gd magnon mode. The appearance only below 100\,mK is presumably because the magnon linewidth becomes narrower than the its coupling to the microwave cavity mode below that temperature \cite{evertsUltrastrongCouplingMicrowave2020}, which is a general condition of a strong coupling between two modes.
The frequency of this mode also appears to decrease with lowering temperature.

\begin{figure*}
    \centering
    \includegraphics[width=\linewidth]{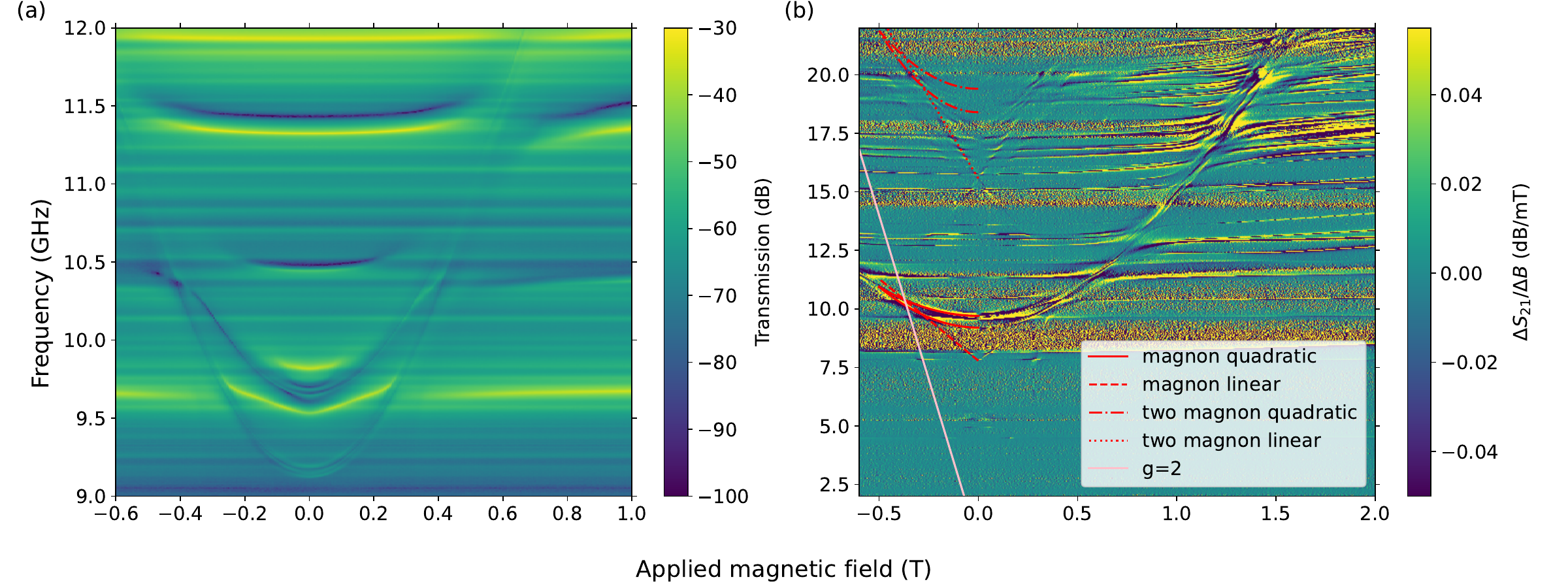}
    \caption{(a) Microwave transmission as a function of applied magnetic field at 60\,mK near the magnons which couple strongly to the cavity modes. They show quadratic field dependence. (b) Difference in transmissions at between adjacent magnetic field points, showing spin excitations which couple weakly to the cavity. The colour scale is harshly clipped to show weak effects. On the negative field side, a number of eye guide lines are drawn which can be reflected to the positive side. ``Two magnon'' lines are drawn at double the single magnon frequencies. }
    \label{fig: S21 field spectra}
\end{figure*}

Microwave transmission as a function of applied magnetic field was measured to determine the magnetic resonance frequencies of Er:GSO (Fig.\,\ref{fig: S21 field spectra}).
Multiple resonances were measured. The two resonances which couple to the cavity most strongly at 9.2 and \qty{9.7}{\GHz} at zero applied field are presumably $k=0$ magnon resonances. 
Their quadratic field dependence is consistent with that of a biaxial antiferromagnet with an applied field along the easy axis \cite{Rezende2019}. 
The two distinct quadratic modes are presumably due to the two gadolinium sites.
In such a biaxial antiferromagnet, the magnon mode which increases quadratically with applied field is expected to be accompanied by a counter-rotating mode at lower frequency which decreases quadratically with applied field.
Weak changes in transmission around \qty{5.2}{\GHz} may be the associated low-frequency modes, however their field dependence was not observed.
Assuming the opposite field dependence to high-frequency magnons, we would expect these modes to soften at about \qty{0.9}{\tesla}, broadly agreeing with the field at which site 1 appears to begin canting.
With an easy $a^*$, intermediate $c$, and hard $b$ axis, we would expect to couple well to the higher frequency magnons with our $b$-oriented AC magnetic field and weakly to the lower frequency magnons. An AC magnetic field along $c$ would then couple well to the lower frequency magnons. We did not test this possibility.

A spin excitation at \qty{7.8}{\GHz} at zero field shows linear field dependence with $g\approx0.5$, but is only observed over a small range due of field. We initially believed it was due to impurity spins, however the appearance of another spin excitation with double the frequency and field dependence supports the notion that this is a magnon mode and that features observed at around \qty{15.6}{\GHz} and above are two-magnon excitations.

Finally, a weak excitation with $g\approx2$ and zero frequency at zero field is seen by its crossings of modes below \qty{10}{\GHz} and from 18 to \qty{20}{\GHz}. We do not know its cause.

\subsection{Photon echo measurements}

We measured two-pulse photon echoes on the main optical transitions of both sites. At \qty{0}{\tesla} and \qty{0.7}{\tesla} applied field, we did not observe echoes with $t_{12}=\qty{5}{\us}$. We therefore assume that the coherence time is low in that field range, possibly limited by low frequency magnons. At \qty{1}{\tesla} and above we could readily observe echoes from both erbium sites. The observed coherence times increased with applied field. We focused on \qty{1.5}{\tesla} and \qty{3}{\tesla} applied field. \qty{1.5}{\tesla} was chosen as it was a low field in this range where all four subsites of sites 1 and 2 were well resolved from each other, and \qty{3}{\tesla} was the highest field we could apply, and where the longest coherence times were measured.

We measured two-pulse photon echoes on each of the four main transitions of sites 1 and 2. We labelled these subsites according to their frequency ordering at \qty{3}{\tesla} applied field (1a--d, 2a--d). Note that subsites 1b and 1c swap order at \qty{1.5}{\tesla}. The amplitudes of these echoes were fitted to a Mims decay $A(t_{12})=A_0\exp[-(2t_{12}/T_{2,M})^{x}]$, where $T_{2,M}$ is the coherence time and $x$ is the stretch factor. Fits to an exponential decay ($x\rightarrow1$, $T_{2,M}\rightarrow T_{2,e}$) did not sufficiently explain echo decay.

\begin{figure*}
    \centering
    \includegraphics[width=\linewidth]{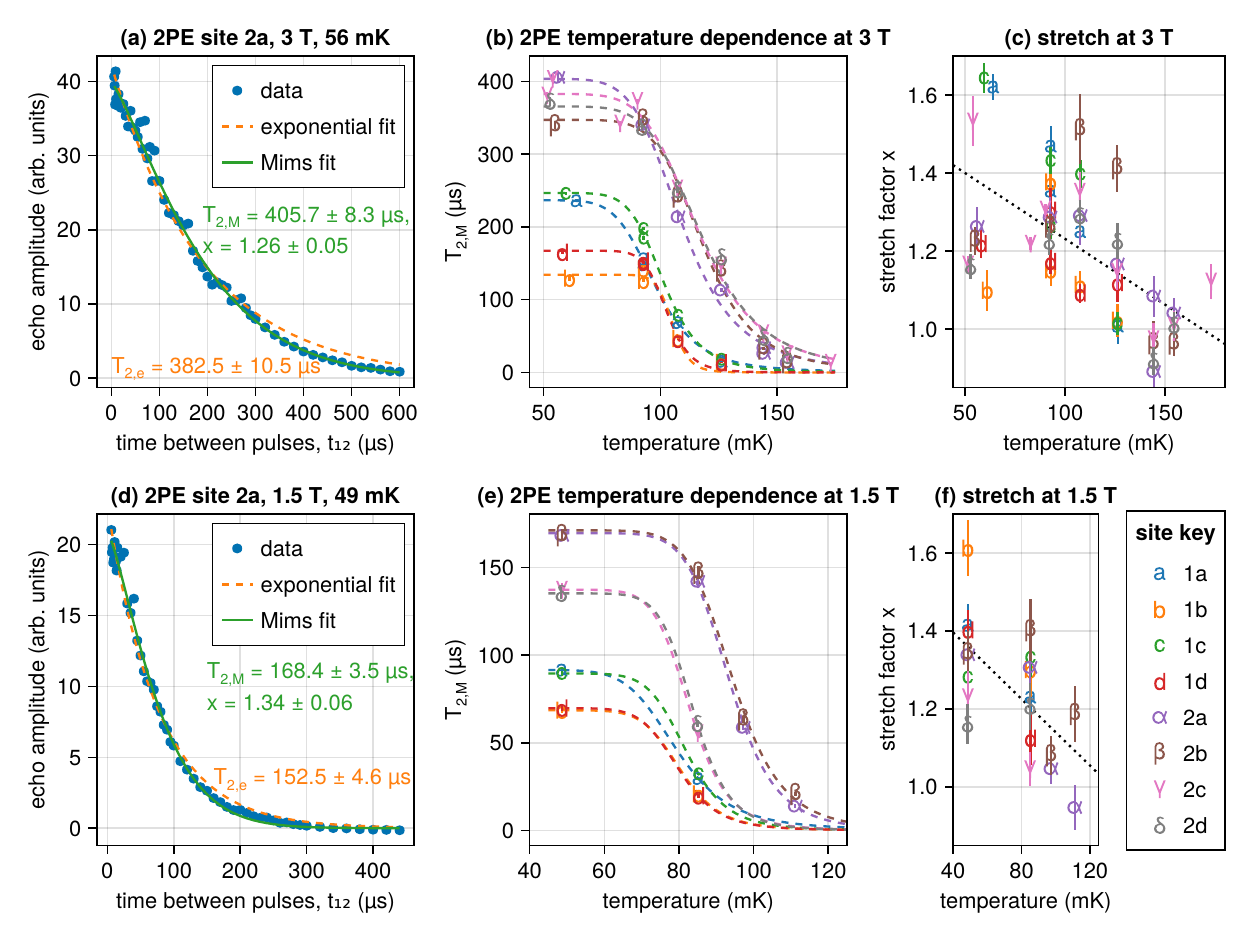}
    \caption{Temperature dependence of coherence time of the different sites at \qty{1.5}{\tesla} and \qty{3}{\tesla} applied field. (a) Measured echo decay at subsite 2a at \qty{3}{\T} applied field along with fits to a Mims decay and exponential decay. (b,c) Fitted coherence times and stretch factors at each subsite at \qty{3}{\T} as a function of temperature recorded by the cavity temperature sensor. The key is given in the bottom right of the figure. Dashed lines in (b) are guides to the eye, assuming the coherence decay is caused by thermally populating a magnon mode \cite{hiraishiLongOpticalCoherence2025c}. The dotted line in (c) is a linear fit of all stretch factors with temperature. (d-f) are the same as (a-c), except at \qty{1.5}{\T} applied field. }
    \label{fig:2PE}
\end{figure*}

At \qty{3}{\tesla} applied field, subsites 2a and 2c exhibit coherence times of \qtylist{406\pm8;404\pm7}{\us}.
The echo decay for subsite 2a is seen in Fig.~\ref{fig:2PE}(a).
As far as we are aware, this makes GSO the host crystal with the fifth-longest optical coherence times ever observed, behind YSO \cite{boettgerMaterialOptimizationEr32003,equallUltraslowOpticalDephasing1994,chiossiOpticalCoherenceSpin2024}, Y\textsubscript{3}Ga\textsubscript{5}O\textsubscript{12} \cite{askaraniLongLivedSolidState2021}, CaWO\textsubscript{4} \cite{tiranovSubsecondSpinLifetimelimited2025}, and Y\textsubscript{2}O\textsubscript{3} \cite{macfarlaneSubkilohertzOpticalLinewidths1981,fukumoriSubkilohertzOpticalHomogeneous2020}, and ahead of YVO\textsubscript{4} \cite{liOpticalSpectroscopyCoherent2020b}, the other material in which a subkilohertz effective homogeneous linewidth, $\Gamma_h=1/(\pi T_{2,M})$, has been measured. 
Notably, GSO is the only magnetically ordered material on this list.

We measured the temperature dependence of coherence times and stretch factors at these two fields, see Fig.~\ref{fig:2PE}. At \qty{3}{\T}, an increase in temperature from $\approx\qty{60}{\milli\K}$ to $\approx\qty{90}{\milli\K}$ only results in a minor degradation of coherence times, whereas at \qty{1.5}{\T} the drop between 50 and \qty{85}{\milli\K} is more severe as would be expected if the limitation were thermally populated magnons with increased frequency at higher fields. The stretch factors generally sit between about 1.6 and 0.9, decreasing as temperature is increased. Linear fits to the temperature dependence of the stretch factors give a $T=0$ intercept of \num{1.57+-0.07} at \qty{3}{\T} and \num{1.57+-0.1} at \qty{1.5}{\T}.
A reduction in stretch factor often occurs as the coherence time becomes similar to the lifetime of the bulk spins perturbing the dopants \cite{mimsPhaseMemoryElectron1968}.

\section{Discussion}

Our results show that measuring the optical transitions of dopants can be an effective means of determining magnetic phases of a host material.
The methods used in this study can be readily expanded upon. In particular, a three axis magnet would allow determination of spin structure without the fortuitous alignment between applied field and spin structure we found in this study. In higher symmetry hosts, the $g$-factor and mean exchange field could be constrained in orientation by site symmetry, eliminating some of the ambiguity seen here for the ordering of site 1.

The high degree of agreement of the absorption spectra of site 2 with the toy model persists at higher temperatures (see Extended Data Fig.~\ref{fig:toysitehot}). As the temperature is increased, the sudden jump in the spectrum smears out to higher and lower field as the excited ground state is populated. Taking the ground doublet to be populated according to the Boltzmann law, the effective spin-\textonehalf{} model gives synthesized spectra that agree well even at higher temperatures. 
This points to the possibility of using Er:GSO as an optical Boltzmann thermometer. 
Er:YSO was explored for this purpose in Ref.~\cite{zemanBoltzmannOpticalThermometry2024}, however their thermometry used two different transitions and required a calibration step of unequal oscillator strength measurement due to the low site symmetry in conjunction with the transition being a mixture of electric- and magnetic-dipole components. These complications are also relevant for Er:GSO, however, the antiferromagnetic spin environments mean we can compare the absorption on the \emph{same} transition from two ions in the same crystal field, but which have differing magnetic environments and therefore different Zeeman splitting. 
From these considerations we believe that Er:GSO could be used to create an optical Boltzmann thermometer at low temperatures, which does not require a temperature calibration step.

Our main motivations for studying these materials are for microwave--optical transduction and for making optical photon memories. In this study we have seen where those two motivations diverge in material selection. For a quantum memory, long coherence times are key, which would favour large geometrically simple antiferromagnetic exchange interactions, such that the material has a high magnon frequencies. 
For microwave--optical transduction, we still want large exchange interactions such that the dopant interacts strongly with the host crystal, however a large variation in sign and magnitude as well as some geometric frustration will pull the magnon frequencies down to the $\sim\qty{5}{\GHz}$ that is ideal for interfacing with superconducting qubits. Here we have seen zero-field magnon frequencies down to $\qty{7.8}{\GHz}$, which are close and plausibly there are lower frequency magnons around \qty{5}{\GHz}. The logical extreme to this comparison is then materials for magnetic refrigeration which require a large density of spins but an ordering temperature that is as cold as possible; the two conditions being possible simultaneously in highly frustrated materials such as gadolinium gallium garnet \cite{petrenkoMagneticFrustrationOrder1999}.

\section{Methods}

\subsection{Sample information and orientation}

We had erbium-doped GSO custom grown with precursor oxides of \qty{99.99}{\percent} purity. The manufacturer measured its doping concentration to be 49\,at.\,ppm by mass spectrometer. It was supplied as a cuboid of size $\qty{5.56}{\mm}\times\qty{5.03}{\mm}\times\qty{5.03}{\mm}$. The \qty{5.56}{\mm} was oriented as the $a^*$ axis, which is perpendicular to the (100) cleavage plane. The other axes of the cuboid were unoriented as purchased. We optically polished the non-(100) faces with lapping film; after a short period of trying to polish the (100) faces, we observed subsurface cleavage a fraction of a millimeter below the sample surface.

We further oriented the sample by finding its optic axes. 
A collimated laser was polarized vertically and focused through the sample. After the sample a horizontal linear polarizer was placed and an interference pattern was observed due to the sample's birefringence. By rotating the sample we could find the centre of the interference pattern and measured the deflection of the light reflected from the sample. Accounting for diffraction at the interface, this told us the direction of an optic axis; the other was found similarly. 
The two optic axes uniquely imply orthogonal axes for the three principal axes of refractive index. Given GSO has positive birefringence, they also show which of these axes has the low, middle and high values of refractive index $n_\alpha < n_\beta < n_\gamma$.
In a monoclinic crystal such as GSO, the symmetry ($b$) axis is one principal axis of refractive index while the other two are perpendicular to $b$, but otherwise unconstrained within the $ac$ plane. 
Only one principal axis was found perpendicular to $a^*$, and therefore we determined it must be the $b$ axis. The $b$ axis thus determined is perpendicular to the plane defined by the two optic axes, showing that light polarized along $b$ has the middle refractive index $n_\beta$.

We estimate that one of the cuboid axes was \qty{9\pm5}{\degree} away from being along the $b$ axis. This cuboid axis was chosen as the light propagation axis. We label the two polarisations of light in the crystal with this propagation direction $\alpha$ and $\gamma$, as they are nearly principal axes of refraction, 
with refractive indices close to $n_\alpha$ and $n_\gamma$.
By measuring the etalon created by the uncoated sample from \qty{194.95}{\THz} to \qty{198.55}{\THz}, we measured the average group index over this range. We found $n_{g,\alpha} = 1.85$, and $n_{g,\gamma} = 1.89$. This measurement was performed cold at $\approx\qty{60}{\milli\K}$. The largest contributor to errors in these values are from the crystal length or propagation angle, but their ratio $n_{g,\gamma}/n_{g,\alpha} = 1.023$ is more precise and close to the value based on known phase indices at \qty{633}{\nm}:  $n_{\gamma}/n_{\alpha} = 1.910/1.871 = 1.021$ \cite{weberHandbookOpticalMaterials2003}. Later, knowledge of the etalon was used to correct for its effects on transmission spectra; because the etalon changed slightly with applied field, this correction is most successful near zero applied field. In the presented results, we used the $\alpha$ polarization, as the measured absorption spectra were very similar, but $\alpha$-polarized light was absorbed a little more.

The sample was mounted in a copper loop--gap microwave cavity shown in Fig.~\ref{fig:sample}(b,c) on the end of a cold finger in the bore of a \qty[number-unit-separator=-]{3}{\T} single-axis superconducting magnet in a dilution refrigerator (Bluefors LD-250). Apiezon N grease was used for thermal contact between the sample and cavity, and the cavity's temperature was monitored by a thermistor (Scientific Instruments RO-600 Group A), which itself was coated in Apiezon N and recessed into the cavity. Measured temperatures were corrected for their magnetic field dependence \cite{ihasLowTemperatureThermometry1998}. Except for our specification of the Néel temperature, we do not include the temperature sensor's accuracy in given temperatures.

To be as unambiguous as possible, we oriented the sample such that an applied magnetic field would be along its $a^*$ axis. The reason was twofold: it was the best oriented axis of our sample, and by applying a field perpendicular to the $b$ axis (or along it), the $g$-factor of all erbium dopants within a given site would be the same.

\subsection{Microwave measurements}

Microwave measurements were made by transmission measurements through the surrounding loop--gap cavity with a vector network analyser (VNA).
Attenuation was used to reduce the effect of thermal noise, and avoid impedance mismatches in the cables \cite{evertsUltrastrongCouplingMicrowave2020}. A detailed setup is given in the Supplement.
We measured transmission through the cavity, $S_{21}$, and investigated temperature and applied magnetic field dependence of magnetic resonances of Er:GSO. For the field dependence, each measurement involved setting a constant magnetic field, measuring a frequency sweep of the VNA and stepping to the next field and repeating. 
The fundamental mode of the loop--gap cavity was at about \qty{2.5}{\GHz}, which has no magnetic field through the sample bore, whereas the lowest-frequency mode that is expected to interact with the sample is at about \qty{4.5}{\GHz}.
In this cavity, an AC magnetic field is generated along the light propagation axis.
The temperature dependence of microwave transmission was measured by repeatedly measuring $S_{21}$ as the refrigerator was cooled from \qty{4}{\kelvin} to its base temperature.

\subsection{Optical absorption spectroscopy}

The optical absorption of the erbium in the sample was determined by optical transmission spectroscopy. A detailed setup is given in the Supplement. Our dilution refrigerator has free-space optical access with four windows on each side of the sample. An integrated telecom laser assembly (ITLA) was scanned across the erbium absorption features. Square pulses were created at about \qty{4}{\kHz} by a pair of acousto-optic modulators (AOMs) with an average power of \qty{2}{\micro\watt} and \qty{50}{\percent} duty cycle. The transmission through the sample as well as a power reference are measured by home-made silicon photodetectors and a pair of lock-in amplifiers. Before each scan, the scan's central frequency was measured by a wavemeter, and during the scan the reflection off of a temperature-stabilized fibre-loop cavity of \qty{94.2}{\MHz} free spectral range were recorded to calibrate and linearize the scan's frequency axis.

\subsection{Effective spin-\textonehalf{} models}

The ground ($g$) and excited ($e$) Kramers doublets are modelled a constant mean-field from their surroundings:
\begin{equation}
\label{eq:toy}
    H_i = \mu_BB_zg_{i,zz}S_{i,z} - \bm{S}_i\cdot\bm{J}_i,
\end{equation}
where $\bm{S}_i$ is a vector of effective spin-\textonehalf{} operators, $B_z$ is the applied magnetic field along the $z=a^*$ axis, $i=g,e$ is the Kramers doublet, and $\bm{J}_i=\langle\sum_{j\in\mathrm{NN}(i)} 2\bm{J}_{ij}\cdot\bm{S}_j\rangle$ is the expectation value of the sum of exchange interactions $\bm{J}_{ij}$ to nearest neighbour (NN) ions $j$ with spin $\bm{S}_j$.
An assumption in this model is that the $z$ axis is a principal axis of the $g$-tensors, i.e. $g_{i,zx}=g_{i,zy}=0$.
We set the mean exchange field to be canted with respect to the applied field axis: $\bm{J}_i = J_i[\sin\theta,0,\cos\theta]$. The angle $\theta$ is then the angle the spin would naturally cant at with respect to the applied field direction.
We used these Hamiltonians to calculate transition strengths assuming only like-to-like transitions as $\propto\lvert{}_g\langle i\vert f \rangle_e\rvert^2$. For an initial state in the ground doublet $i$ and a final state in the excited doublet $f$. Oscillator strength on like-to-unlike transitions in this model then arises due to angles between the ground- and excited-doublet quantisation axes \cite{fernandez-gonzalvoCoherentFrequencyUpconversion2015}.

The main lines' frequency dependence gives the g-factor difference, whereas weak absorption features observed between \qty{0.3}{\T} and \qty{0.5}{\T} give their sum such that $g_{e,zz}\equiv g_{e,a^*}=7.3$ and $g_{g,zz}\equiv g_{g,a^*}=8.7$. The ground state exchange splitting is found from the field the observed transition changes: $J_g=-\mu_Bg_{g,a^*}B_\mathrm{mf}/h$. $B_\mathrm{mf}$ is the applied magnetic field that cancels the Zeeman splitting of the ground state due to exchange. In the data, we see four sublattices with $B_\mathrm{mf}=\pm\qty{0.505}{\T},\pm\qty{0.54}{\T}$. The difference in exchange parameters is seen as the frequency separation between parallel-line absorptions above $B_\mathrm{mf}$ from which we find that $J_e$ shares the same sign as $J_g$, but is \qty{20}{\GHz} lower in magnitude. 
A low canting angle of $\theta\approx\qty{0.5}{\degree}$ gives oscillator strength to the like-to-unlike transitions we observed, but a canting above \qty{1}{\degree} is not plausible as it would cause the lines to curve in the vicinity of $B_\mathrm{mf}$.

Some qualitative agreement for site 1 is obtained with $g_{g,zz} = 4=g_{g,zz}$, $J_g=\qty{-25}{\GHz}$, $J_e=\qty{-5}{\GHz}$, and a canting angle of $\theta=\qty{20}{\degree}$.
Better agreement for site 1 at the expense of an extra parameter can be found with a mean field along the $a^*$ axis, but with anisotropic $g$-tensors where $a^*$ is not a principal axis. 
If we assume that site 1 is CN9, and assume its approximate $C_{3v}$ symmetry, we can take the $g$-tensors to be
\begin{align}
    \bm{g}_{i,C_{3v}} &{}\approx R_x(\phi)\begin{bmatrix}
        g_{i,\perp} & 0 & 0 \\
        0 & g_{i,\perp} & 0 \\
        0 & 0 & g_{i,\parallel}
    \end{bmatrix}R_x^\intercal(\phi) 
    \\
    &{}\approx \frac12\begin{bmatrix}
        2g_{i,\perp} & 0 & 0 \\
        0 & g_{i,\parallel}+g_{i,\perp} & g_{i,\parallel}-g_{i,\perp} \\
        0 & g_{i,\parallel}-g_{i,\perp} & g_{i,\parallel}+g_{i,\perp}
    \end{bmatrix},
\end{align}
where $\phi\sim\qty{45}{\degree}$ is the angle between the $a^*$ and approximate $C_{3}$ axis, and $R_x(\phi)$ is a rotation by that angle about the $x$ axis. Note that the coordinate system for this $g$-tensor is arbitrary, but with the $z$-axis coinciding with $a^*$. The Hamiltonian is then
\begin{equation}
\label{eq:C3v model}
    H_i = \mu_BB_z\left[\frac{g_{i,\parallel}+g_{i,\perp}}2S_{i,z}+\frac{g_{i,\parallel}-g_{i,\perp}}2S_{i,y}\right] - S_{i,z}J_i,
\end{equation}
where we have removed the angle between applied field and mean exchange field. To produce Fig.~\ref{fig:coldspectra}(c), we have used the parameters $g_{g,\perp} = 4.6$, $g_{g,\parallel} = 3.6$, $g_{e,\perp} = 5.6$, $g_{e,\parallel} = 2.6$, $J_g=\qty{-23}{\GHz}$, $J_e=\qty{-5}{\GHz}.$

\subsection{Photon echoes}

We measured coherence times of the optical transitions with a two-pulse photon echo experiment. A detailed setup is given in the Supplement.
The pair of AOMs were used to create square pulses of peak power about \qty{1.85}{\milli\watt} in the pulse sequence $\pi/2-\pi-\text{echo}$. The light sent through the sample was coupled into an optical fibre and mixed with a local oscillator from the laser in a 50/50 splitter. The splitters outputs were detected on a balanced heterodyne detector (Thorlabs PDB835C-AC). The resulting heterodyne beat was at \qty{10.7}{\MHz} as a result of the AOM frequencies of $+\qty{75}{\MHz}$ and $-\qty{85.7}{\MHz}$. This beat was filtered and amplified by an amplifier with a fast recovery time such that weak echoes could be measured next to the strong signals from driving laser pulses. The resulting \qty{10.7}{\MHz} was digitized and recorded to extract echo amplitudes. The laser was free running in these experiments, which caused some echoes to have reduced amplitude due to frequency fluctuations. Therefore, at each $t_{12}$, we took 100 echoes. Using the highest echo amplitudes with an unstabilized laser is known to accurately reproduce the true decay that would be obtained with a stabilized laser \cite{stricklandLaserFrequencyStabilization2000}.

\subsection{Photoluminescence assignment of sites}

Previous studies had not identified the ${}^4I_{13/2}(Y_1)$ levels of both sites in the crystal, for example Ref.~\cite{dingTemperatureEffectEmission2011} only observed site 1, and the frequency separation of about \qty{100}{\giga\hertz} could be within their measurement error. Therefore, to assign the close-in-frequency transitions we performed an emission-resolved photoluminescence measurement, where the $Z_1\rightarrow Y_1$ transitions were excited, but spontaneous emission from $Y_1\rightarrow Z_2$ was measured.

On the fibre output for echo measurements, we added a manually tunable fibre bandpass filter with \qty{1}{\nm} bandwidth (Agiltron FOTF-025121333), followed by a superconducting nanowire single photon detector counted by a time tagger. Pulses of power \qty{10.6}{\micro\watt} and \qty{60}{\ms} duration were separated by \qty{60}{\ms} of dead time, during which a histogram of photon counts was created. The filter was first set to \qty{30}{\per\cm} (corresponding to the first crystal field splitting of site 1) lower in frequency than the average measured frequency of both sites. The laser was set to excite each site and a histogram of photoluminescence was built up over \qty{100}{\s}. One site showed a strong photoluminescence signal with the \qty{30}{\per\cm} detuning and was assigned to site 1; the other site showed a strong signal with \qty{57}{\per\cm} detuning and was assigned to site 2. 
After assignment, the filter position and pump frequency were tweaked to optimize count rate and the measurement was repeated to find the lifetime, $T_1$, of each site, as well as $Z_2$ energy.
Data from the photoluminescence (PL) measurement is given in Extended Data Fig.~\ref{fig:PL}.
From these measurements the lifetimes of the ${}^4I_{13/2}$ states at \qty{90}{\milli\kelvin} were determined to be \qty{8.87\pm0.01}{\ms} for site 1 and \qty{5.14\pm0.01}{\ms} for site 2, and $Z_2$ energies were \qty{29+-2}{\per\cm} for site 1 and \qty{58+-2}{\per\cm} for site 2. The shorter lifetime for site 2 is perhaps the reason it was not observed in Ref.~\cite{dingTemperatureEffectEmission2011}. Interestingly site 1 has a lifetime much closer to its room temperature value (\qty{8.4}{\ms}) than its \qty{13}{\K} value (\qty{13.2}{\ms}) \cite{dingTemperatureEffectEmission2011}. 
The lower site 2 lifetime is one reason we do not assume the site assignments of \textcite{camargoSpectroscopicCharacteristicsEr32002}.

\subsection{Temperature control}

Measurements were made at elevated temperatures in various ways. We measured field dependence at fixed temperatures by heating the mixing chamber plate and servoing its temperature. Due to the experiment's distance from the heater, we did not servo its temperature directly, but found it stabilized to the servoed temperature with an offset. Above about \qty{1}{\K}, it was necessary to reduce forced evaporation from the condensed helium by turning the refrigeration cycle's turbo pump off. A cooldown measurement was made with \qty{3}{\tesla} applied field by condensing the mixture at that field. To determine the Néel temperature as accurately as possible in a temperature range that is difficult for a dilution refrigerator, we condensed only about \qty{20}{\percent} of the helium mixture for normal operation and operated the dilution refrigerator similar to a 1-K pot. The temperature of the mixing chamber plate was servoed and gradually reduced as measurements were taken.

\section*{Acknowledgements}

This work was supported by Quantum Technologies Aotearoa, a research programme of Te Whai Ao -- the Dodd Walls Centre, funded by the New Zealand Ministry of Business Innovation and Employment through International Science Partnerships, contract number UOO2347.

Fig.~\ref{fig:sample}(a) was made using VESTA \cite{mommaVESTA3Threedimensional2011}.

\renewcommand{\figurename}{Extended Data Fig.}
\setcounter{figure}{0}

\begin{figure*}
    \centering
    \includegraphics[width=\linewidth]{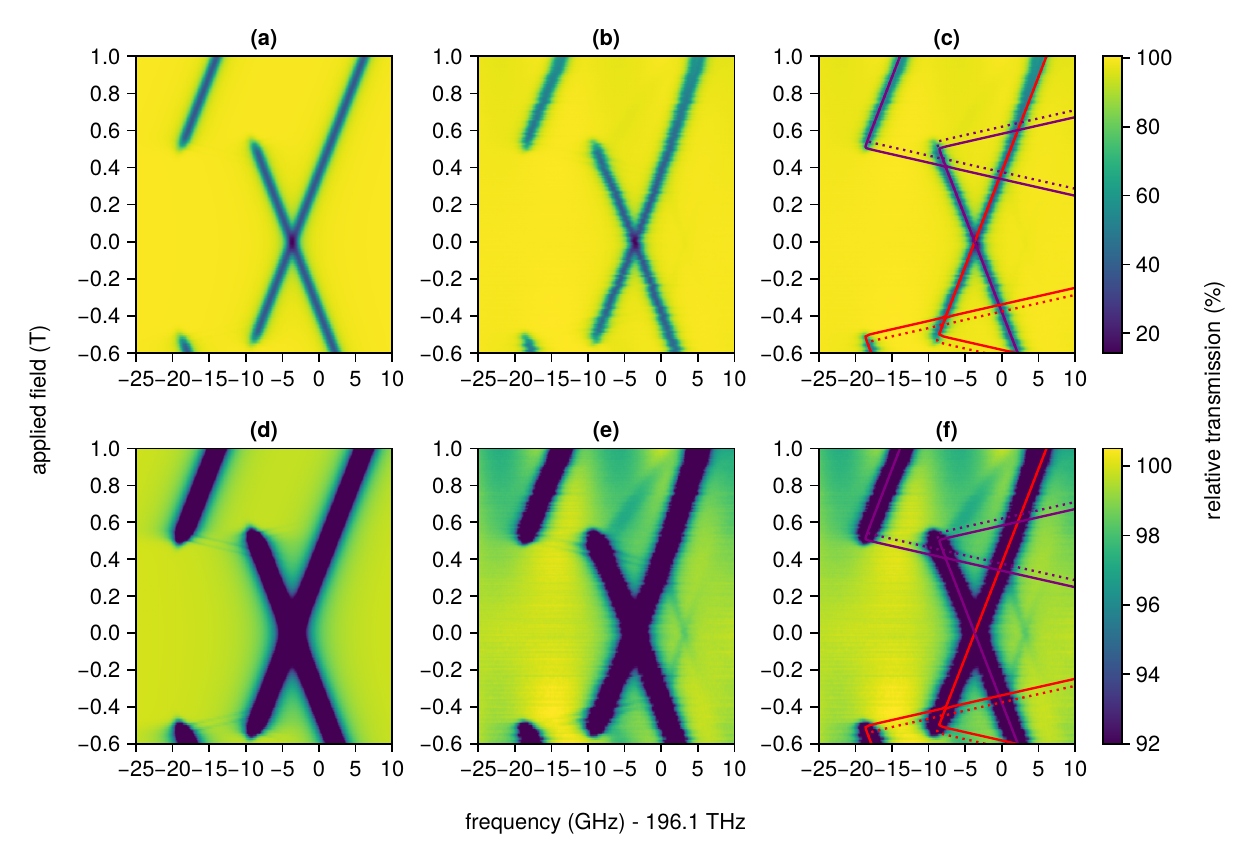}
    \caption{Comparison of synthetic spectra from the effective spin-\textonehalf{} model to real spectra for site 2 at the base temperature of the dilution refrigerator. Parameters given in main text. The assumed misalignment was \qty{0.5}{\degree}. (a,d) Synthetic spectra. (b,e) Experimental data. (c,f) Experimental data with calculated transitions from the ground state of the model overlaid. Spin-up (spin-down) sublattices at zero field are shown as purple (red) lines. Dotted lines are the sublattices with higher $|B_\mathrm{mf}|$. The difference between (a--c) and (d--f) is just the colour scale.}
    \label{fig:toysite2}
\end{figure*}

\begin{figure*}
    \centering
    \includegraphics[width=\linewidth]{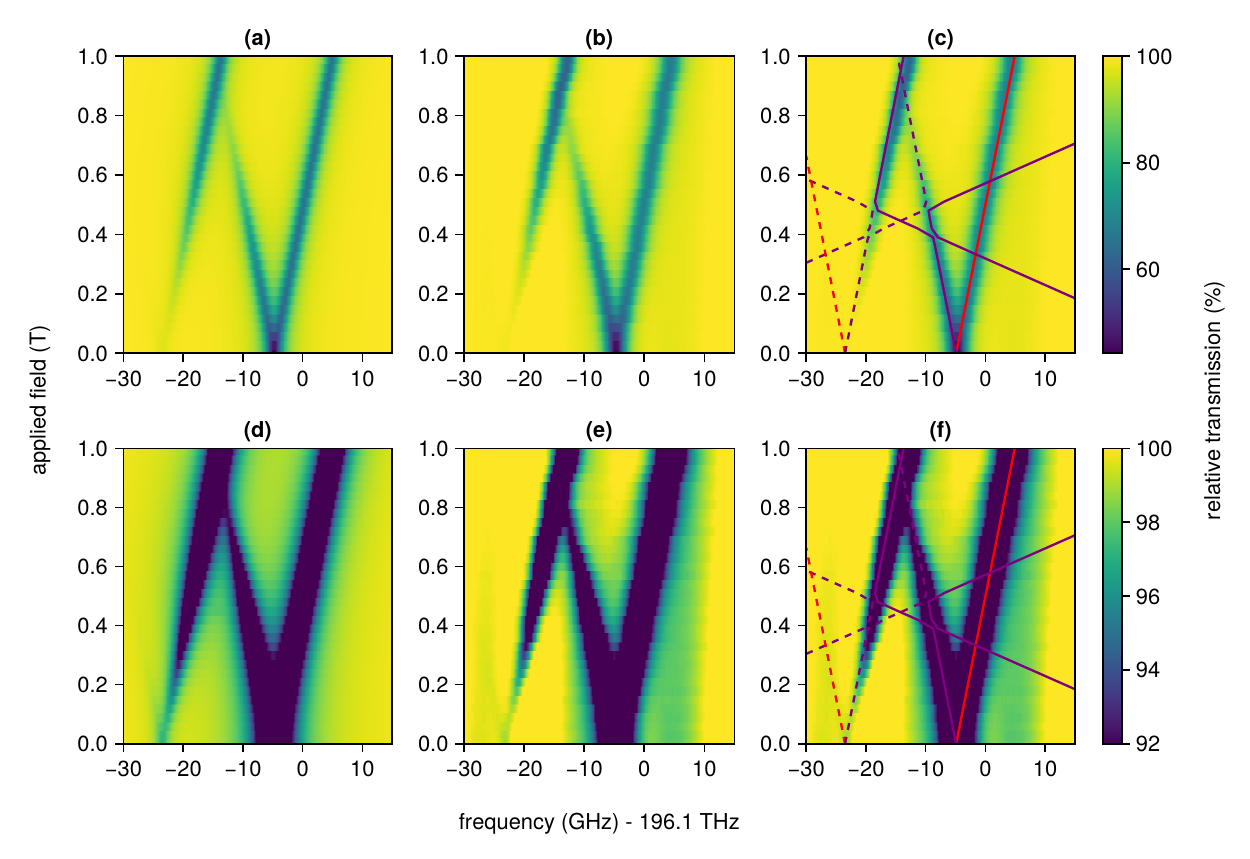}
    \caption{Comparison of synthetic spectra from the effective spin-\textonehalf{} model to real spectra for site 2 at \qty{830\pm10}{\milli\kelvin}. (a,d) Synthetic spectra. (b,e) Experimental data. (c,f) Experimental data with calculated transitions from the ground state of the model overlaid. Spin-up (spin-down) sublattices at zero field are shown as purple (red) lines. Solid (dashed) lines are transitions from the ground (first excited) state. The difference between (a--c) and (d--f) is just the colour scale. The differences to the cold spectra in Extended Data Fig.~\ref{fig:toysite2} are that the assumed temperature is higher, the exchange splitting is reduced to \qty{93}{\percent} of the cold value to account for reduction in mean field, and a Lorentzian FWHM of \qty{2}{\GHz} is used.}
    \label{fig:toysitehot}
\end{figure*}

\begin{figure*}
    \centering
    \includegraphics[width=\linewidth]{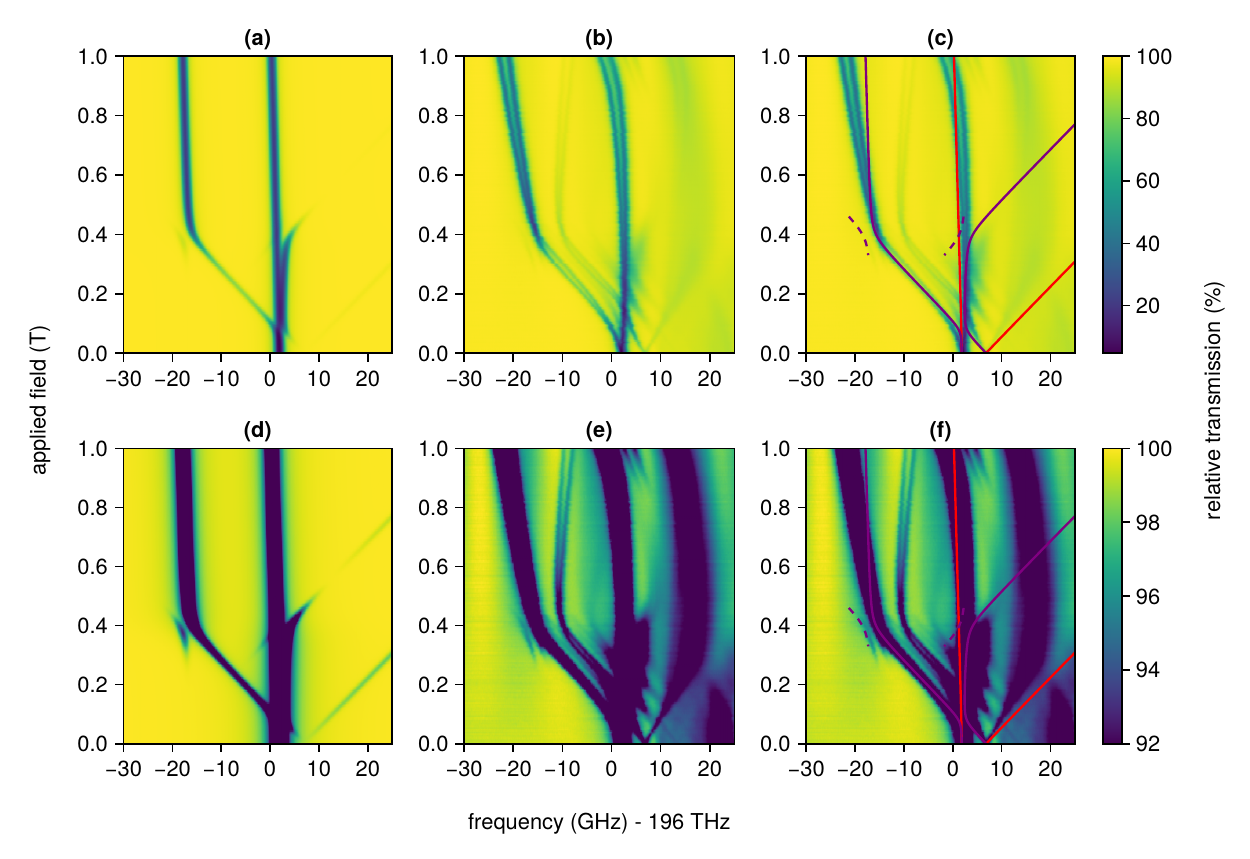}
    \caption{Comparison of synthetic spectra from the $C_{3v}$ effective spin-\textonehalf{} model of Eq.~\eqref{eq:C3v model} to real spectra for site 1 at the base temperature of the dilution refrigerator. Parameters: $g_{g,\perp} = 4.6$, $g_{g,\parallel} = 3.6$, $g_{e,\perp} = 5.6$, $g_{e,\parallel} = 2.6$, $J_g=\qty{-23}{\GHz}$, $J_e=\qty{-5}{\GHz}.$ (a,d) Synthetic spectra. (b,e) Experimental data. (c,f) Experimental data with calculated transitions from the ground state of the model overlaid. Spin-up (spin-down) sublattices at zero field are shown as purple (red) lines. Dashed lines are thermal lines where the ground state and first excited state are within \qty{5}{\GHz}. The difference between (a--c) and (d--f) is just the colour scale.}
    \label{fig:toysite1C3v}
\end{figure*}

\begin{figure*}
    \centering
    \includegraphics[width=\linewidth]{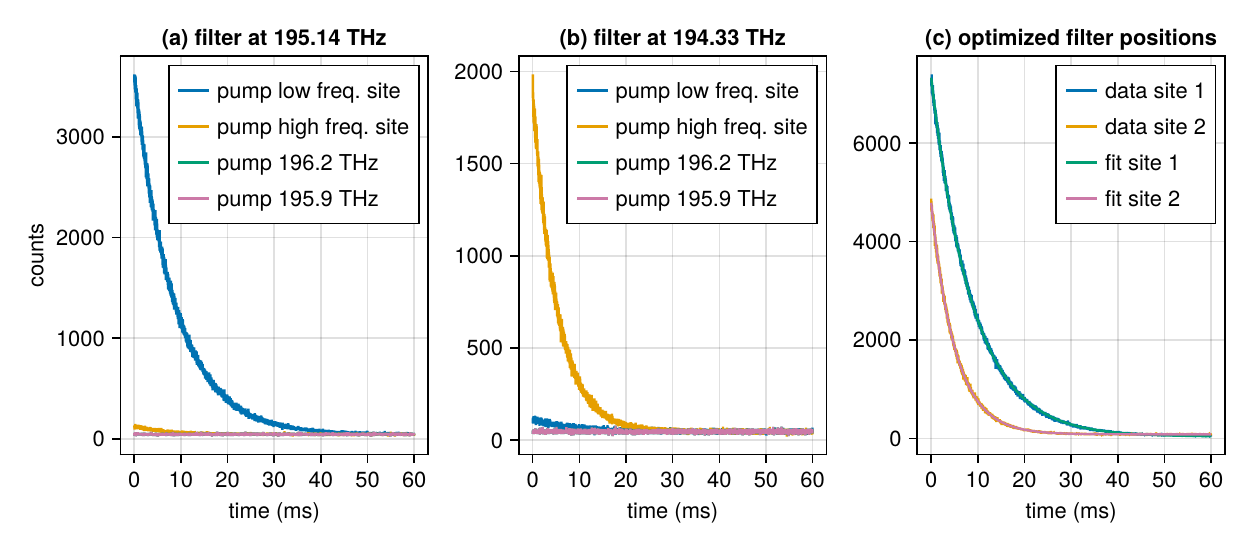}
    \caption{Filtered photoluminescence (PL) measurement. Decay from $Y_1$ to $Z_2$ is measured. (a) The fiber filter is set to \qty{195.14}{\THz}, and PL is seen for the lower frequency erbium site. (b) The fiber filter is set to \qty{195.33}{\THz}, and PL is seen for the higher frequency erbium site. Low frequency site pump${}=\qty{196.0027}{\THz}$; high frequency site pump${}=\qty{196.0973}{\THz}$. In both cases pumping at frequencies above and below the absorption features are control measurements. (c) For pumping each site, the fiber filter and pump frequency are tweaked to obtain the best PL signal. From these measurements, the optical lifetimes are fit.}
    \label{fig:PL}
\end{figure*}

\clearpage

\title{Supplementary information for: \texorpdfstring{\\}{} Elucidating magnetic structure with optical dopants: erbium-doped Gd\texorpdfstring{\textsubscript{2}}{₂}SiO\texorpdfstring{\textsubscript{5}}{₅}}

\maketitle

\renewcommand{\figurename}{Fig.}
\setcounter{section}{0}
\setcounter{equation}{0}
\setcounter{figure}{0}
\setcounter{table}{0}
\makeatletter
\renewcommand{\thesection}{S\arabic{section}}
\renewcommand{\theequation}{S\arabic{equation}}
\renewcommand{\thefigure}{S\arabic{figure}}
\renewcommand{\thetable}{S\arabic{table}}
\renewcommand{\bibnumfmt}[1]{[S#1]}
\renewcommand{\citenumfont}[1]{S#1}

\renewcommand{\thesection}{S\arabic{section}}
\renewcommand{\theequation}{S\arabic{equation}}
\renewcommand{\thefigure}{S\arabic{figure}}
\renewcommand{\thetable}{S\arabic{table}}
\renewcommand{\bibnumfmt}[1]{[S#1]}
\renewcommand{\citenumfont}[1]{S#1}

\onecolumngrid

\vspace{-1\baselineskip}
\section{Detailed setup}

\begin{figure*}
    \centering
    \includegraphics[width=\linewidth]{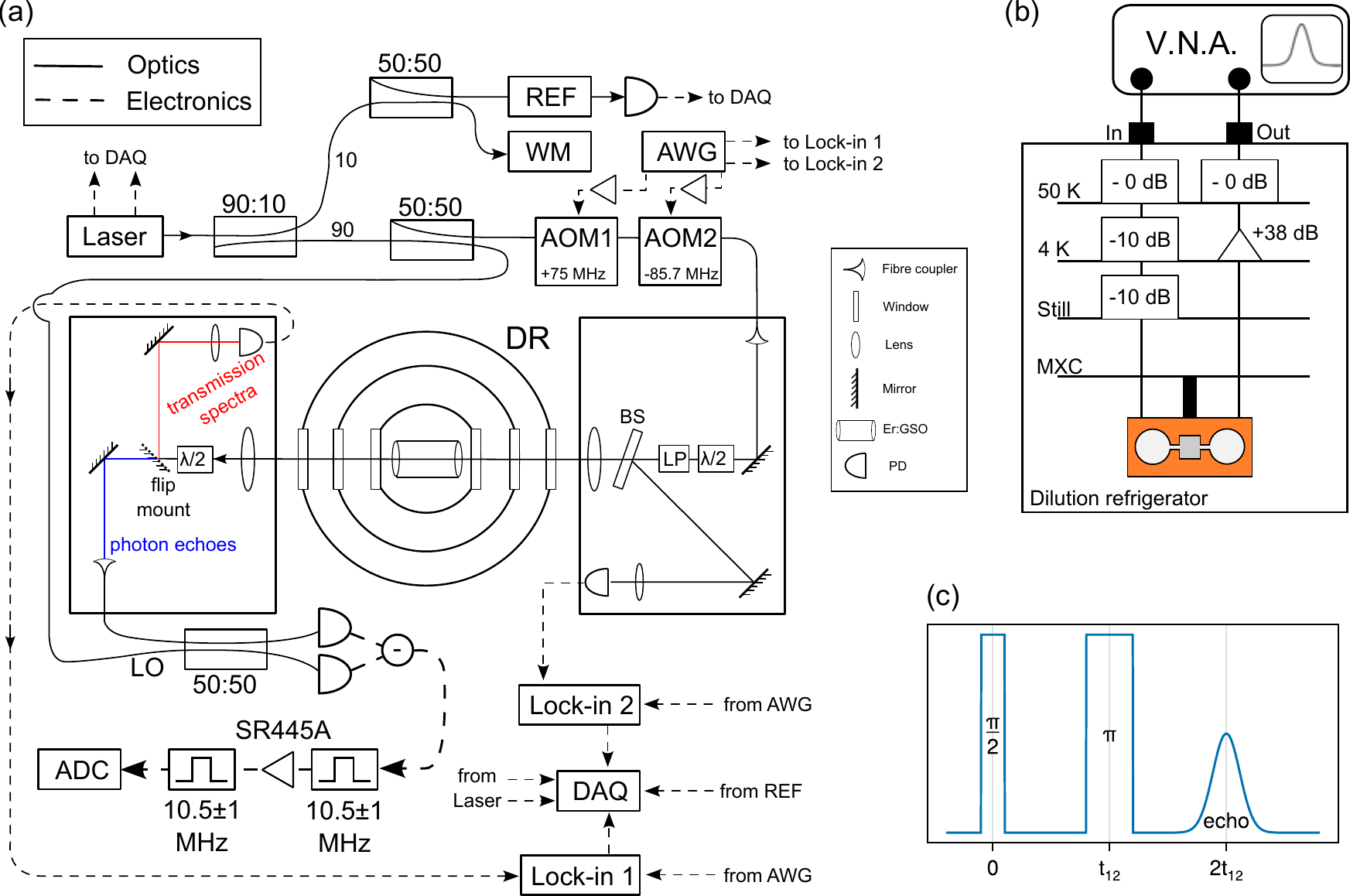}
    \caption{(a) Setup for optical experiments. ADC, analog-to-digital converter; AOM, acousto-optic modulator; AWG, arbitrary waveform generator; DAQ, data aquisition; DR, dilution refrigerator; LO, local oscillator; LP, linear polarizer; PD, photodetector; REF, reference cavity; WM, wavemeter. (b) Setup of microwave measurements, showing microwave cables, attenuators, and amplifiers inside the dilution refrigerator. MXC, mixing chamber plate; V.N.A., vector network analyzer. (c) Pulse sequence for two-pulse photon echoes.}
    \label{fig:setup}
\end{figure*}

A detailed setup for optical and microwave experiments is given in Fig.~\ref{fig:setup}. In the optical setup after the dilution refrigerator there are two paths for the light that travelled through the sample depending on the detection scheme. The path taken is controlled by a mirror on a flip mount.

\end{document}